\documentclass[aip,apl,twocolumn,reprint,superscriptaddress,floatfix]{revtex4-1}
\usepackage{float}

\usepackage{graphicx}
\usepackage{dcolumn}
\usepackage{bm,bbm}
\usepackage{pstricks,pst-plot}     
\usepackage{latexsym}
\usepackage{mathrsfs,amsmath,amssymb}
\usepackage{color}
\usepackage{balance}
\definecolor{MidnightBlue}{RGB}{0,0,160}
\usepackage[colorlinks=true, citecolor=MidnightBlue,linkcolor=black,urlcolor=MidnightBlue, breaklinks=true]{hyperref}
\usepackage{booktabs}

\begin{document}


\title{Photocarrier extraction in GaAsSb/GaAsN type-II QW superlattice solar cells}

\author{U. Aeberhard}
\email{u.aeberhard@fz-juelich.de}

\affiliation{IEK-5 Photovoltaik, Forschungszentrum J\"ulich, D-52425 J\"ulich,
Germany}

\author{A. Gonzalo}

\author{J. M. Ulloa}

\affiliation{Institute for Systems based on Optoelectronics and Microtechnology (ISOM), Universidad Politécnica de Madrid, Avda. Complutense 30, 28040 Madrid, Spain}

\date{\today}

\begin{abstract}
Photocarrier transport and extraction in GaAsSb/GaAsN type-II quantum well superlattices are investigated by means of inelastic quantum transport calculations based on the non-equilibrium Green's function formalism. Evaluation of the local density of states and of the spectral current flow enables the identification of different regimes for carrier localization, transport, and extraction as a function of configurational parameters. These include the number of periods, the thicknesses of the individual layers in one period, the built-in electric field, and the temperature of operation. The results for the carrier extraction efficiency are related to experimental data for different symmetric GaAsSb/GaAsN type-II  quantum well superlattice  solar cell devices and provide a qualitative explanation for the experimentally observed dependence of photovoltaic device performance on period thickness.  
\end{abstract}

\maketitle

Multi-junction solar cells enable photovoltaic solar energy conversion efficiencies beyond 46\% \cite{green:18_pip} due to enhanced utilization of the solar spectrum and the reduction of thermalization losses \cite{marti:04,green:06}. Maximization of the efficiency requires suitable combination of materials with specific band gap energies. Among the most promising candidates are 4-junction devices based on (Al)InGaP and (In)GaAs  as top cells, where the challenge is to find a third cell with band gap energy of 1-1.15 eV that is lattice matched to (In)GaAs and the Ge substrate \cite{toprasertpong:16}. Recently, strain-balanced GaAsSb/GaAsN quantum well superlattice (QWSL) structures have been proposed as high-quality subcell materials with tunable band gap close to the target range \cite{gonzalo:17,gonzalo:18_spie}. In addition to the absence of clustering \cite{reyes:12} due to spatial separation of N and Sb, the type-II SL (SL-II) offers largely independent control of conduction and valence band edges via the individual adjustment of the N and Sb content, respectively \cite{utrilla:14}. 
\begin{figure}[t]
	\begin{center}
	\includegraphics[width=0.4\textwidth]{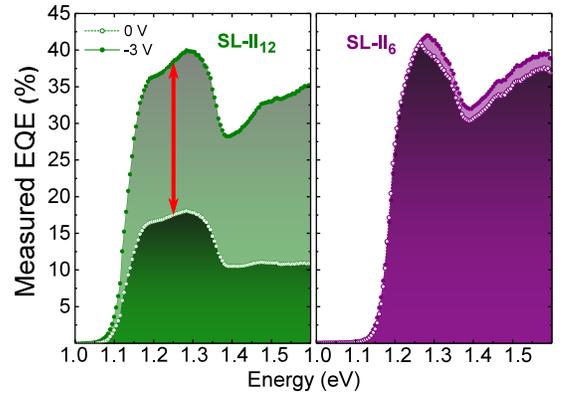}
	\caption{Measured external quantum efficiency for GaAsSb/GaAsN QWSL solar cells with period thicknesses of 12 nm (SL-II$_{12}$) and 6 nm (SL-II$_{6}$), respectively. Comparison of the data at short circuit ($V=0$ V) and at large reverse bias ($V=-3$ V), where all of the photogenerated carriers are extracted, reveals a massive deterioration of the extraction efficiency for the thick period, which points at configuration-dependent transport issues.\label{fig:eqe_exp}}
	\end{center}
\end{figure}
The tunability of the effective band gap by variation of the SL period thickness $d_{\textrm{per}}$ was confirmed by photoreflectance measurements for $d_{\textrm{per}}\in\{3,6,12,20\}$ nm \cite{gonzalo:17}, showing effective band gap energies in good agreement with theoretical predictions from SL dispersion calculations based on effective masses and band offsets derived from the double band-anticrossing (2-BAC) model \cite{lin:08,lin:13}. In order to move towards low effective band gaps, larger period thicknesses are favored in principle. However, as displayed in Fig.~\ref{fig:eqe_exp}, the external quantum efficiency (EQE) experiments revealed serious performance issues related to incomplete carrier extraction at short circuit conditions for $d_{\textrm{per}}=12$~nm (SL-II$_{12}$) as compared to the situation for $d_{\textrm{per}}=6$~nm (SL-II$_{6}$), where the coincidence of the EQE at short circuit and at large reverse bias points at unit extraction efficiency \cite{gonzalo:17}. Hence, the SL period thickness has critical impact on the photovoltaic device performance, which requires careful analysis in order to identify optimum configurations. 

\begin{figure}[t]
	\includegraphics[width=0.5\textwidth]{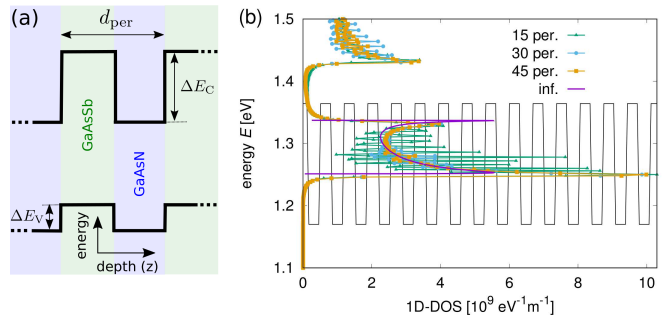}
	\caption{(a) Schematic layer structure and band profile of the GaAsSb/GaAsN type-II SL considered in this work. (b) Evolution of the 1D density of states (at $\mathbf{k}_{\parallel}=0$) in the first miniband of the SL-II$_{6}$ configuration with increasing number of periods. For comparison, the SL potential and the DOS of the infinite (perfectly periodic) SL are shown as well. While the miniband width is converged already at 15 periods, finite-size effects remain as small features in the DOS up to very large period numbers.\label{fig:elstruct}}
\end{figure}

Here, we perform such an analysis by means of rigorous simulation of photocarrier transport in GaAsSb/GaAsN QWSL structures with varying thicknesses of the GaAsSb and GaAsN layers. Our approach relies on steady-state quantum-kinetic theory in the shape of the non-equilibrium Green's function formalism (NEGF) as adopted for the description of photovoltaic device operation in nanostructure-based solar cell devices \cite{ae:jcel_11,ae:jmr_18}. As in previous applications of NEGF to the simulation of QWSL solar cells \cite{ae:nrl_11,ae:jpe_14}, the electronic structure is modeled in terms of a simple effective mass Hamiltonian describing decoupled conduction and valence bands. Planar geometry is assumed, with the in-plane spatial coordinates Fourier-transformed to transverse momentum $\mathbf{k}_{\parallel}$. In the present case, the effective masses of the bulk ternaries and the band offsets defining the SL heterostructure potential along the growth direction $z$ are obtained from the 2-BAC model. While this amounts to a crude simplification of the complex electronic structure of dilute nitrides and does not allow for quantitatively accurate computations in an extended spectral range, it provides a reasonable qualitative description of the states occupied via optical transitions at photon energies close to the effective band gap, as considered here. Table \ref{tab:param} displays the parameters for N content $x_{\textrm{N}}=1.2\%$ and Sb concentration $y_{\textrm{Sb}}=3.25\%$, as experimentally determined for the samples of Ref.~\onlinecite{gonzalo:17}. This configuration with low N and Sb contents is nearly lattice-matched to GaAs and exhibits very low levels of strain in the constituent layers; the effects of strain are therefore neglected in the model. The corresponding band offsets given by the 2-BAC model are \mbox{$\Delta E_{\textrm{C}}=194$ meV} and \mbox{$\Delta E_{\textrm{V}}=50$ meV} [cf. Fig.~\ref{fig:elstruct} (a)].  Any other material parameters required for the quantification of the interaction of carriers with photons and phonons, i.e., the parametrization of the respective interaction Hamiltonians, are chosen as of bulk GaAs. 

\begin{table}[b!]
	\caption{Bulk parameters for constituent materials used in the simulations as derived from the double band-anticrossing model.\cite{lin:08,lin:13}}
	\vspace{1mm}
	\begin{tabular}{r|c|c}
		\hline
		\hline
		&GaAsSb$_{0.0325}$&GaAsN$_{0.012}$\\
		\hline
		el. affinity $\chi$ (eV)&4.076&4.270\\
		band gap $E_{\textrm{g}}$ (eV)&1.364&1.220\\
		el. effective mass $m_{e}^{*}$/$m_{0}$&0.067&0.098\\
		hl. effective mass $m_{h}^{*}$/$m_{0}$&0.516&0.500\\
		\hline
		\hline
	\end{tabular}
	\label{tab:param}
\end{table}
 
Conventionally, the properties of SL structures are analyzed based on models assuming perfect periodicity and flat band conditions. The appropriateness of both assumptions for the finite systems under consideration will be assessed in the following. Figure \ref{fig:elstruct}(b) displays the one-dimensional density of electron states (at vanishing transverse momentum, $\mathbf{k}_{\parallel}=0$) of the GaAsSb/GaAsN type-II SL with $d_{\textrm{per}}=6$ nm obtained by integrating the local density of states (LDOS) from the Green's functions over the entire finite structure, $\mathscr{D}_{\mathrm{1D}}(E)=-L_{z}^{-1}\int dz \Im G^{R}(\mathbf{0},z,z,E)/\pi$. The 1D density of states (1D-DOS) is shown for increasing number of SL periods, revealing a fast convergence of the miniband width, but features related to the finite size of the structure remain up to very large systems. For comparison, we also plot the 1D-DOS of the perfectly periodic SL, which is evaluated by $\mathscr{D}_{\mathrm{1D}}(E)=2\{\partial_{q_z}\varepsilon_{n}(q_z)\}^{-1}/\pi|_{q_z(\varepsilon_{n}=E)}$ using the longitudinal SL dispersion $\varepsilon_n(q_{z})$ for $n=1$ as obtained from solving the Schr\"odinger equation for the same effective mass Hamiltonian, but with periodic boundary conditions. There is a close agreement between finite and infinite system DOS for the band width, however, some discrepancies remain in shape and asymmetry of the DOS inside the band.

\begin{figure}[t!]
	\includegraphics[width=0.45\textwidth]{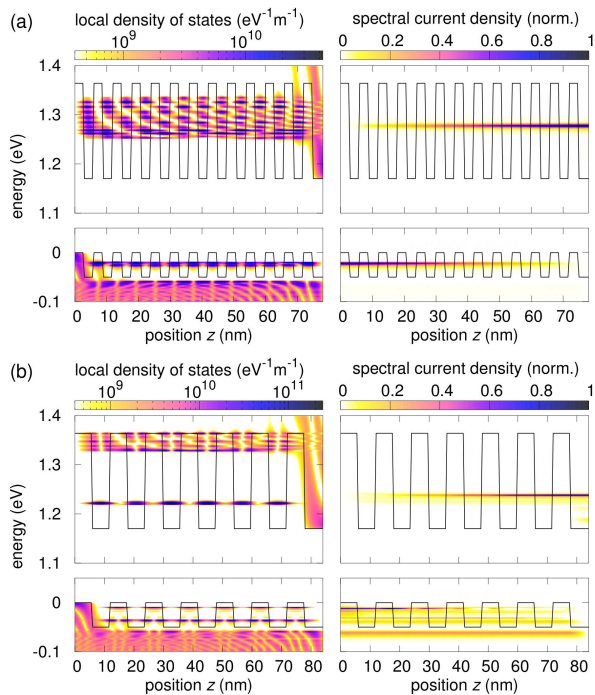}
	\caption{Local density of states (for $\mathbf{k}_{\parallel}=0$) and (normalized) spectral current flow under monochromatic illumination for (a) SL-II$_6$ (at $E_{\gamma}=1.3$ eV) and (b) SL-II$_{12}$ (at $E_{\gamma}=1.25$ eV) configurations in the absence of built-in fields and inelastic scattering. In both cases, the states are delocalized over the entire SL and extraction proceeds quasi-ballistically at the energy of generation, with no indication of a thickness-dependent change in the qualitative behavior. \label{fig:flatband}}
\end{figure}

In order to investigate localization and transport of photogenerated charge carriers in finite SL structures with different period thicknesses, we evaluate the (1D) LDOS - corresponding to the integrand of the formula for the 1D-DOS - and the spectral current density $j(E)=\mp(2e\hbar/m_{0})\lim_{z'\rightarrow z}\big(\partial_{z}-\partial_{z'}\big)\int d{\mathbf{k}_{\parallel}}G^{\lessgtr}(\mathbf{k}_{\parallel},z,z',E)/(2\pi)^3$, where the upper (lower) sign is for electrons (holes), for the quasi-ballistic case, i.e., in the presence of elastic scattering only.  In Fig.~\ref{fig:flatband}(a), the result is displayed for a SL-II$_{6}$ structure with 12 periods and carrier-selective contacts as required for photovoltaic device operation \cite{wuerfel:book_09}. The LDOS shows clear miniband formation and wave function delocalization over the entire structure for the electrons. In the case of the holes, coupling of periods is weaker, but still visible. The quasi-ballistic current under monochromatic illumination with photon energy $E_{\gamma}=1.3$ eV flows within the miniband at the energy where carriers are generated. The picture does not change significantly in Fig.~\ref{fig:flatband}(b) for a 6-period SL-II$_{12}$ structure, except for stronger wave function localization and the appearance of a second confined miniband for both electrons and holes. However, a qualitative change in the carrier extraction as experimentally observed can not be deducted from this quasi-ballistic flat-band scenario. In the next step, the localization and transport of photocarriers in type-II GaAsSb/GaAsN SL is therefore investigated in a more realistic picture including a finite built-in electric field and considering inelastic scattering of electrons and holes with optical phonons.

\begin{figure*}[t]
	\begin{center}
		\includegraphics[width=0.95\textwidth]{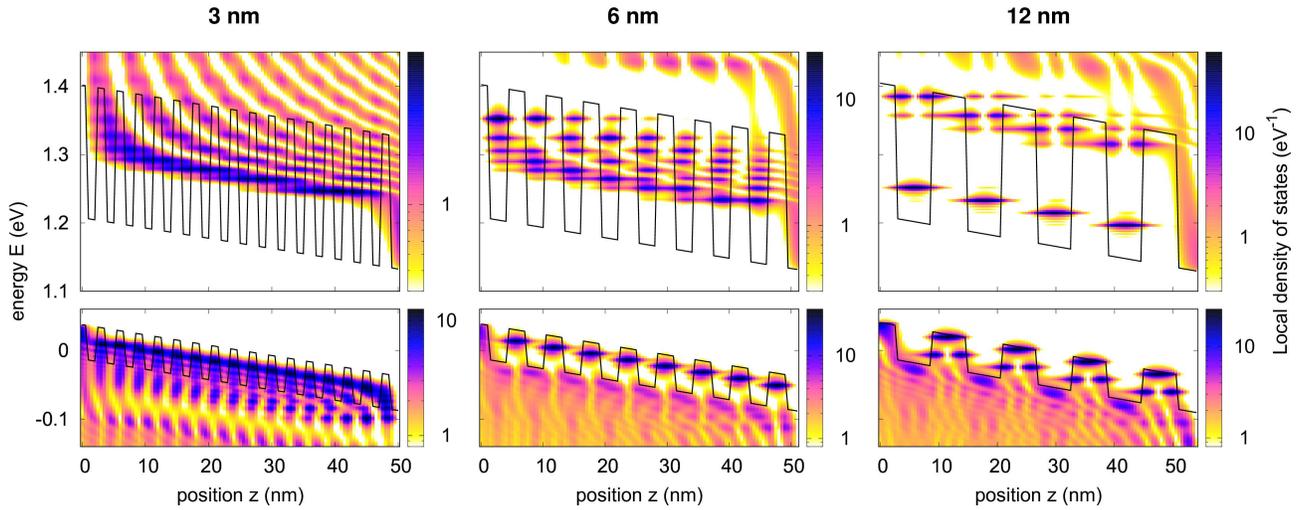}
		\caption{1D-LDOS for SL structures of increasing period thickness, exhibiting the pronounced evolution of localization features: at $d_{\mathrm{per}}=3$ nm, both electron and and hole states are delocalized over many periods; at $d_{\mathrm{per}}=6$ nm, holes have become localized, while electrons are still delocalized; at $d_{\mathrm{per}}=12$ nm, the lowest states of both carrier species are localized. \label{fig:ldos_3612}}
	\end{center}
\end{figure*}

\begin{figure*}[t]
	\begin{center}
		\includegraphics[width=0.95\textwidth]{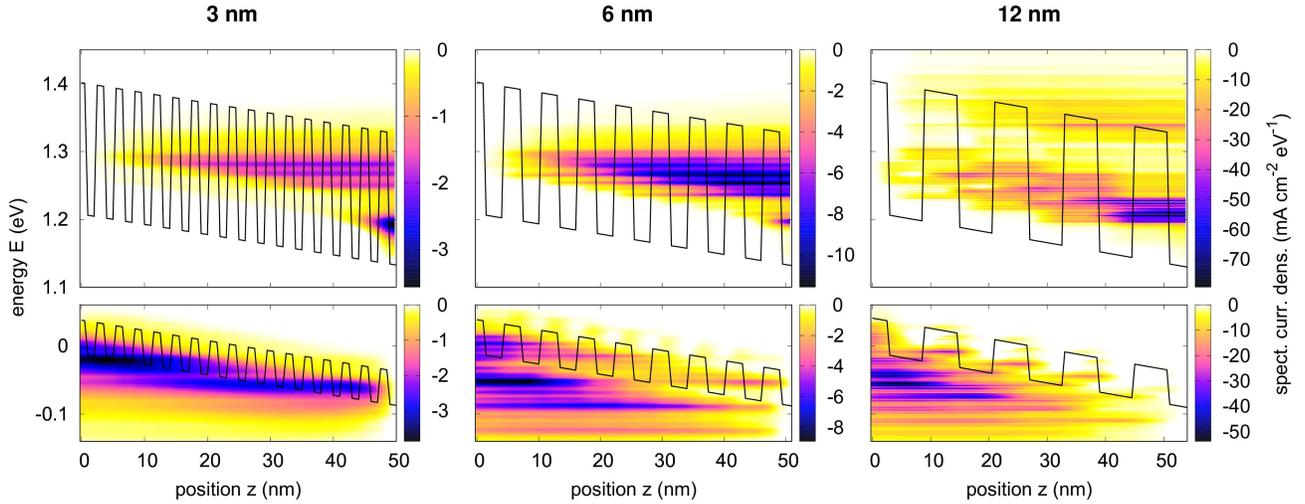}
		\caption{The thickness-dependent carrier localization is reflected in the spectral current density (at $E_{\gamma}=1.25$ eV), revealing different transport regimes in dependence of period thickness: at $d_{\mathrm{per}}=3$ nm, carrier extraction proceeds quasi-ballistically for both electrons and holes; at $d_{\mathrm{per}}=6$ nm, the quasi-ballistic extraction is restricted to the electrons, while for the holes, a significant part of the current is thermally activated through the absorption of optical phonons; at $d_{\mathrm{per}}=12$ nm, there is a hot carrier current component from thermal escape to the continuum also for the electrons, and the tunneling proceeds in sequential fashion with pronounced inter-period carrier relaxation.   \label{fig:spectcurr_3612}}
	\end{center}
\end{figure*}

\begin{figure}[t!]
	\begin{center}
	\includegraphics[width=0.48\textwidth]{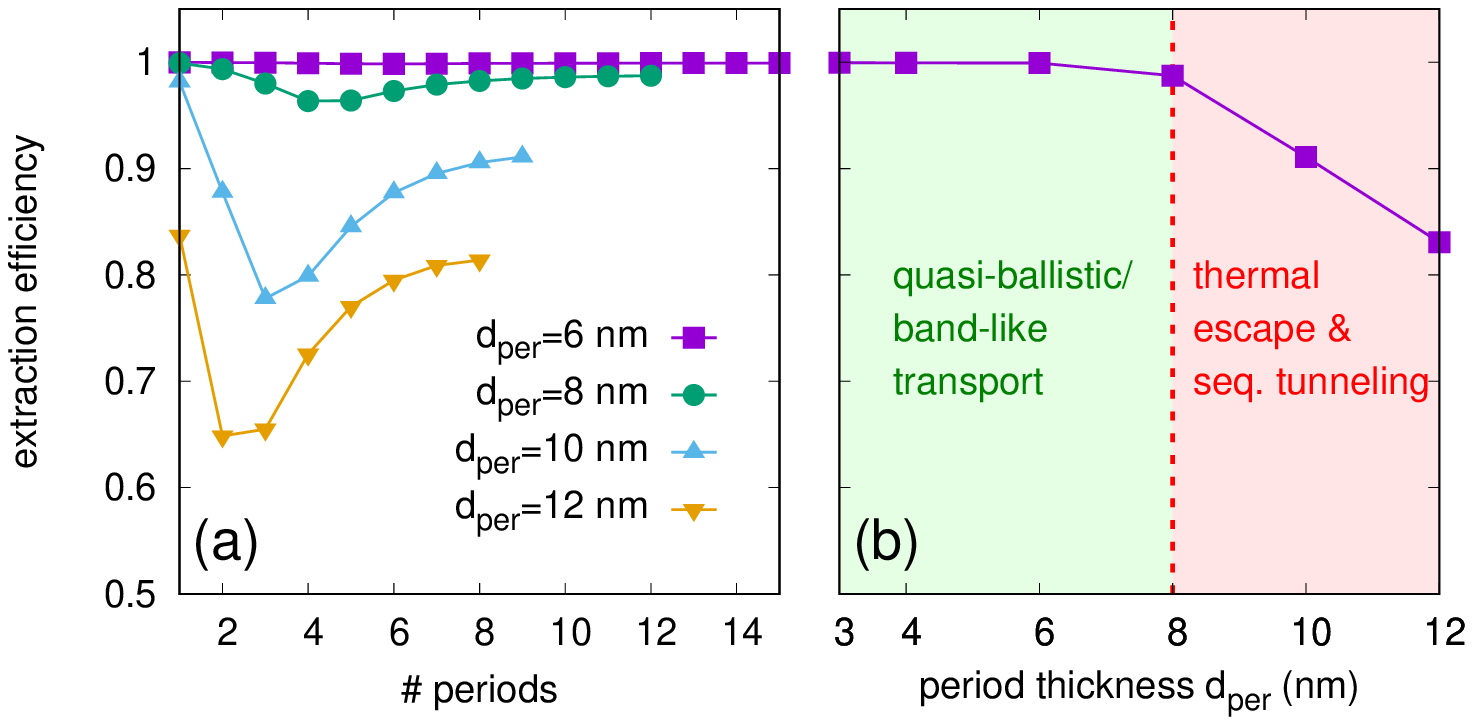}
	\caption{(a) Convergence of the carrier extraction efficiency with the number of SL periods for different period thicknesses at $E_{\gamma}=1.25$ eV and $\tau\sim 10$ ps. Due to stronger localization, convergence is faster for thicker periods. (b) Converged carrier extraction efficiency as a function of period thickness: there is a pronounced drop around $d_{\mathrm{per}}=8$ nm, which can be attributed to a transition in the transport regime from quasi-ballistic or band-like to sequential tunneling with a sizable hot carrier current component due to phonon-assisted thermal escape.  \label{fig:extracteff_perdep}}
	\end{center}
\end{figure}

Figure \ref{fig:ldos_3612} shows the electron and hole 1D-LDOS at a built-in field of 15 kV/cm -- obtained from a drift-diffusion-Poisson simulation of the full experimental $p$-$i$-$n$ structures described in Ref.~\onlinecite{gonzalo:17} -- and at short-circuit conditions ($V_{\textrm{bias}}=0$ V). For $d_{\textrm{per}}=3$ nm, both electrons and holes are delocalized over multiple periods. At $d_{\textrm{per}}=6$ nm, the hole states are confined to a single period, while the electron wave functions are still delocalized. Finally, at $d_{\textrm{per}}=12$ nm, both electron and hole ground states are fully localized. The thickness-dependent localization behavior is reflected in the transport mechanism for charge carriers, as shown in Fig.~\ref{fig:spectcurr_3612} displaying the spectral photocarrier flow, again for monochromatic illumination with $E_{\gamma}=1.25$ eV. At $d_{\textrm{per}}=3$ nm, photocarrier extraction is fast and proceeds quasi-ballistically for both electrons and holes, irrespective of the presence of inelastic electron-phonon scattering. At $d_{\textrm{per}}=6$ nm, electron extraction proceeds still quasi-ballistically, while escape of holes by tunneling is no longer fast, which leads to the appearance of a hot carrier current component that is thermally activated by the absorption of optical phonons. At  $d_{\textrm{per}}=12$ nm, also the electron flow has developed a significant thermal component, while the tunneling of electrons proceeds in a sequential fashion. This transition in the transport mechanism from quasi-ballistic extraction to sequential tunneling with thermal component has dramatic consequences for the carrier extraction efficiency, as will be shown next.

Formally, the carrier extraction efficiency is defined as the ratio of carrier extraction and generation rates, which can be expressed as $\eta_{\textrm{ext}}=J_{\textrm{sc}}/J_{\textrm{gen}}$, where $J_{\textrm{sc}}$ is the short circuit current and $J_{\textrm{gen}}=q\int dz ~G(z)$ the generation current obtained from the local generation rate $G$. For mesoscopic extension of the active device region as considered here, incomplete extraction is expected only for very low carrier lifetime, as induced by the presence of fast non-radiative recombination channels. As the description of non-radiative recombination as part of a complete NEGF treatment of photovoltaic devices is still at an early stage \cite{ae:mrs_12}, and thus only radiative recombination is included here, the low-lifetime regime is made accessible by boosting the rate for spontaneous emission. The lifetime $\tau$ is then approximated via the relation $\mathcal{R}_{\textrm{rad}}=\delta\rho/\tau$, where $\delta\rho$ is the photogenerated excess carrier density. In order to study the impact of the period thickness on the photocarrier extraction efficiency, the latter is evaluated for values of $d_{\textrm{per}}$ from 3 nm to 12 nm and at an effective carrier lifetime of \mbox{$\tau\sim 10$ ps} \footnote{The qualitative extraction picture does not depend on the exact value of $\tau$, as long as 1/$\tau$ is comparable to the rate of carrier escape to the contact reservoirs.}. To exclude boundary effects due to low band gap contact layers, a buffer period is inserted on both sides. To obtain results that are relevant for large period number devices (as investigated experimentally), the extraction efficiency is converged with respect to the number of SL periods. The convergence shown in Fig.~\ref{fig:extracteff_perdep}(a) for $E_{\gamma}=1.25$ eV reflects the increasing degree of decoupling of neighboring SL periods with growing period thickness. Figure \ref{fig:extracteff_perdep}(b) displaying the converged values as a function of period thickness reveals a dramatic drop in carrier extraction efficiency around $d_{\textrm{per}}=8$ nm, which can now be attributed to the transition in the photocarrier transport regime from quasi-ballistic to sequential tunneling accompanied by phonon-mediated escape. 

\begin{figure}[b!]
	\begin{center}
	\includegraphics[width=0.5\textwidth]{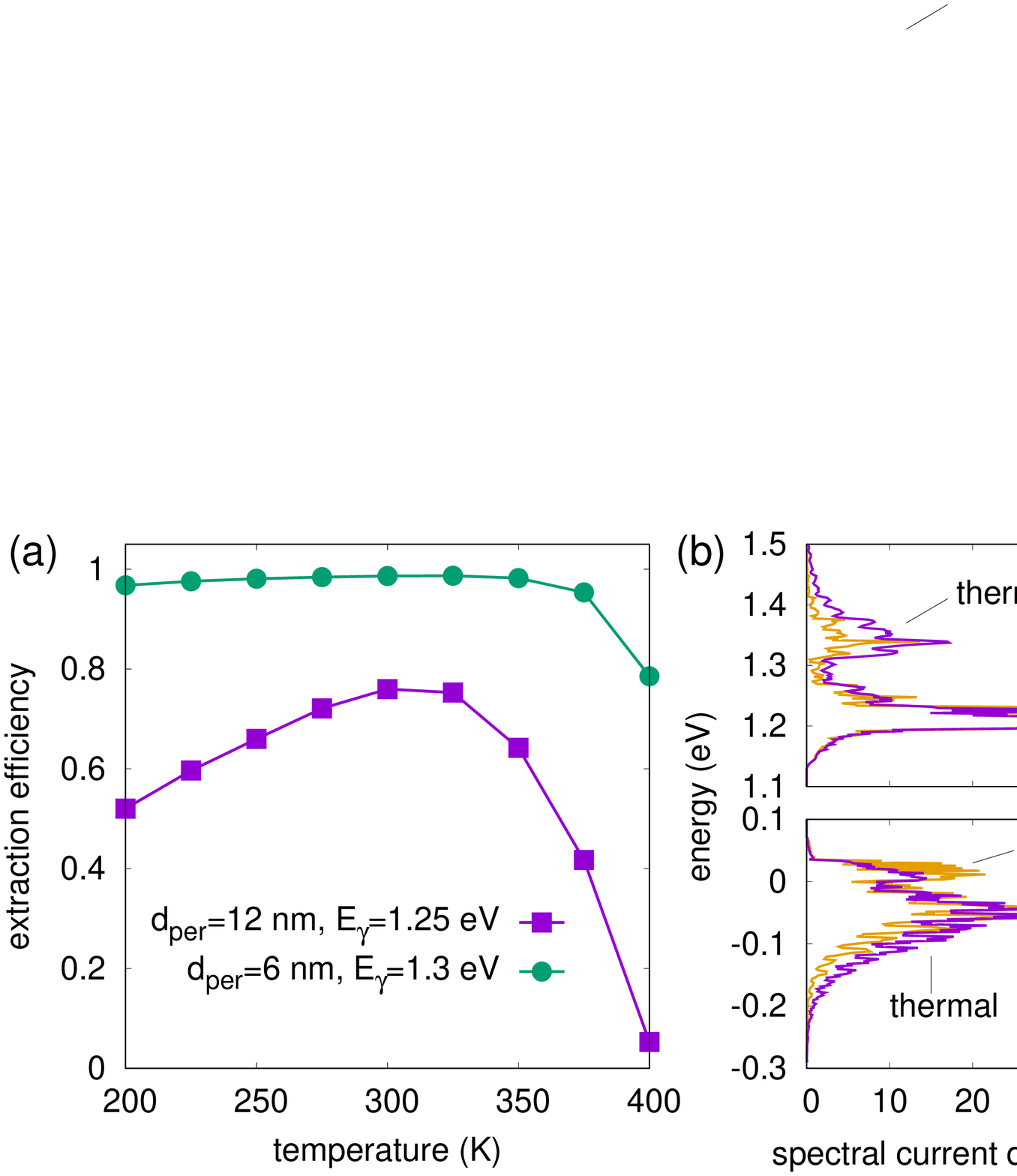}
	\caption{(a) Temperature dependence of the carrier extraction efficiency for $d_{\mathrm{per}}=6$ nm at $E_{\gamma}=1.3$ eV and \mbox{$d_{\mathrm{per}}=12$ nm} at $E_{\gamma}=1.25$ eV. Due to the relevance of thermal escape for carrier extraction in the thick period SL, the temperature dependence is very pronounced in this case, while in the thin period SL, extraction is almost insensitive to temperature in the range relevant for device operation. (b) The current spectrum at the majority contacts for $d_{\mathrm{per}}=12$ nm displays clearly distinguishable contributions from tunneling and thermal escape, with tunneling dominating electron transport, while hole transport is to a large extent thermally activated. With increasing temperature (250/350 K: light/dark color), current weight is transferred from the tunneling component to the thermally activated component. \label{fig:tempdep}}
	\end{center}
\end{figure}

Since thermally activated escape has been identified as an important extraction mechanism at larger period thicknesses, it is instructive to consider the temperature dependence of the extraction efficiency. The results shown in Fig.~\ref{fig:tempdep}(a) confirm the previously established picture in that close to room temperature, transport at $d_{\textrm{per}}=6$ nm does not rely on thermally activated processes, while such processes play a crucial role at $d_{\textrm{per}}=12$ nm. The decrease of extraction efficiency at higher temperatures can be explained by the larger population of phonon modes resulting in increased scattering-induced carrier localization. Figure \ref{fig:tempdep}(b) displaying the current spectrum at the contacts for T=250 K (light color) and T=350 K (dark color), respectively, shows the transfer of current weight from the tunneling to thermal component with increasing temperature. At both temperatures shown, the tunneling component dominates for electrons and the thermal component for holes.

\begin{figure}[t]
	\begin{center}
		\includegraphics[width=0.48\textwidth]{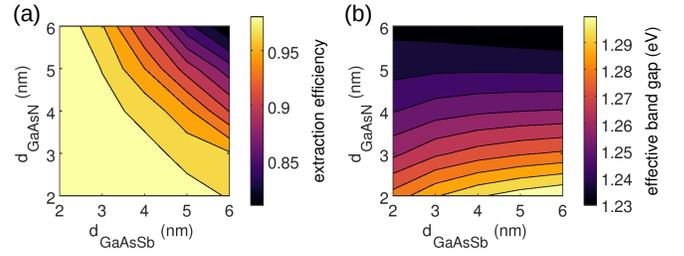}
		\caption{(a) Photocarrier extraction efficiency at \mbox{$E_{\gamma}=1.3$ eV} for different thicknesses of the GaAsSb and GaAsN layers in one period. (b) Effective band gap of the corresponding QWSL structures. In combination, periods with thin GaAsSb and thick GaAsN layers are identified as configurations promising enhanced photovoltaic performance. \label{fig:extracteff_bg_2d}}
	\end{center}
\end{figure}

In principle, the thicknesses of GaAsSb and GaAsN layers can be varied independently (under the constraint of observing the strain-balancing condition), which offers a wider parameter space for the design of optimized superlattice solar cell structures. Application of our model and simulation approach to the parameter space spanned by \mbox{$d_{\textrm{GaAsSb/GaAsN}}\in\{2,3,4,5,6\}$ nm} identifies configurations with thin GaAsSb and thick GaAsN layers as ideal combination exhibiting high photocarrier extraction efficiency at low effective band gap values, as can be inferred from Figs.~\ref{fig:extracteff_bg_2d}(a) and \ref{fig:extracteff_bg_2d}(b). 

In conclusion, we performed quantum-kinetic simulations of photocarrier generation, transport and recombination in GaAsSb/GaAsN type-II SL solar cells which shed light on the carrier extraction mechanism in dependence of the geometrical configuration, built-in field and temperature. The experimental observation of a dramatic drop in carrier extraction efficiency from 6~nm to 12~nm period thickness is explained by the transition in the dominant electron transport regime from quasi-ballistic or band-like transport to sequential tunneling accompanied by phonon-assisted thermal escape. Improved carrier extraction efficiency at similar effective band gap as in the symmetric 12-nm period structure is predicted for asymmetric configurations with reduced GaAsSb layer thickness. 

The approach demonstrated here may thus serve as a valuable tool in the computational design of superlattice solar cells. In future work, it could for instance be used to explore compositions with effective band gaps closer to the target value for the application in multijunction devices. For a comprehensive assessment of the impact of SL configuration on the photovoltaic device performance, in addition to the intrinsinc absorber under uniform field, the contact regions subject to sizable band bending will need to be considered as well, as they might affect carrier extraction, similar as observed, e.g., in simulations of quantum dot SL absorbers \cite{berbezier:15} or ultra-thin solar cell architectures \cite{ae:apl_16}.  

This work benefited from COST action MP1406 -- MultiscaleSolar. UA acknowledges support through the European Commission Horizon 2020 project No. 676629, as well as compute time granted on the supercomputer JURECA at the J\"ulich Supercomputing Centre (JSC). AG and JMU acknowledge Spanish MINECO through project MAT2016-77491-C2-1-R.

\balance

%


\begin{thebibliography}{19}%
\makeatletter
\providecommand \@ifxundefined [1]{%
 \@ifx{#1\undefined}
}%
\providecommand \@ifnum [1]{%
 \ifnum #1\expandafter \@firstoftwo
 \else \expandafter \@secondoftwo
 \fi
}%
\providecommand \@ifx [1]{%
 \ifx #1\expandafter \@firstoftwo
 \else \expandafter \@secondoftwo
 \fi
}%
\providecommand \natexlab [1]{#1}%
\providecommand \enquote  [1]{``#1''}%
\providecommand \bibnamefont  [1]{#1}%
\providecommand \bibfnamefont [1]{#1}%
\providecommand \citenamefont [1]{#1}%
\providecommand \href@noop [0]{\@secondoftwo}%
\providecommand \href [0]{\begingroup \@sanitize@url \@href}%
\providecommand \@href[1]{\@@startlink{#1}\@@href}%
\providecommand \@@href[1]{\endgroup#1\@@endlink}%
\providecommand \@sanitize@url [0]{\catcode `\\12\catcode `\$12\catcode
  `\&12\catcode `\#12\catcode `\^12\catcode `\_12\catcode `\%12\relax}%
\providecommand \@@startlink[1]{}%
\providecommand \@@endlink[0]{}%
\providecommand \url  [0]{\begingroup\@sanitize@url \@url }%
\providecommand \@url [1]{\endgroup\@href {#1}{\urlprefix }}%
\providecommand \urlprefix  [0]{URL }%
\providecommand \Eprint [0]{\href }%
\providecommand \doibase [0]{http://dx.doi.org/}%
\providecommand \selectlanguage [0]{\@gobble}%
\providecommand \bibinfo  [0]{\@secondoftwo}%
\providecommand \bibfield  [0]{\@secondoftwo}%
\providecommand \translation [1]{[#1]}%
\providecommand \BibitemOpen [0]{}%
\providecommand \bibitemStop [0]{}%
\providecommand \bibitemNoStop [0]{.\EOS\space}%
\providecommand \EOS [0]{\spacefactor3000\relax}%
\providecommand \BibitemShut  [1]{\csname bibitem#1\endcsname}%
\let\auto@bib@innerbib\@empty
\bibitem [{\citenamefont {Green}\ \emph {et~al.}(2018)\citenamefont {Green},
  \citenamefont {Hishikawa}, \citenamefont {Dunlop}, \citenamefont {Levi},
  \citenamefont {Hohl-Ebinger},\ and\ \citenamefont
  {Ho-Baillie}}]{green:18_pip}%
  \BibitemOpen
  \bibfield  {author} {\bibinfo {author} {\bibfnamefont {M.~A.}\ \bibnamefont
  {Green}}, \bibinfo {author} {\bibfnamefont {Y.}~\bibnamefont {Hishikawa}},
  \bibinfo {author} {\bibfnamefont {E.~D.}\ \bibnamefont {Dunlop}}, \bibinfo
  {author} {\bibfnamefont {D.~H.}\ \bibnamefont {Levi}}, \bibinfo {author}
  {\bibfnamefont {J.}~\bibnamefont {Hohl-Ebinger}}, \ and\ \bibinfo {author}
  {\bibfnamefont {A.~W.}\ \bibnamefont {Ho-Baillie}},\ }\href {\doibase
  10.1002/pip.2978} {\bibfield  {journal} {\bibinfo  {journal} {Prog.
  Photovolt: Res. Appl.}\ }\textbf {\bibinfo {volume} {26}},\ \bibinfo
  {pages} {3} (\bibinfo {year} {2018})}\BibitemShut {NoStop}%
\bibitem [{\citenamefont {Mart\'i}(2004)}]{marti:04}%
  \BibitemOpen
  \bibinfo {editor} {\bibfnamefont {A.}~\bibnamefont {Mart\'i}},\ ed.,\ \href
  {http://wwwzb.fz-juelich.de/contentenrichment/inhaltsverzeichnisse/2012/0750309059.pdf}
  {\emph {\bibinfo {title} {Next generation photovoltaics : high efficiency
  through full spectrum utilization}}},\ Series in optics and optoelectronics\
  (\bibinfo  {publisher} {IOP},\ \bibinfo {address} {Bristol},\ \bibinfo {year}
  {2004})\BibitemShut {NoStop}%
\bibitem [{\citenamefont {Green}(2006)}]{green:06}%
  \BibitemOpen
  \bibfield  {author} {\bibinfo {author} {\bibfnamefont {M.~A.}\ \bibnamefont
  {Green}},\ }\href {http://dx.doi.org/10.1007/b137807} {\emph {\bibinfo
  {title} {Third Generation Photovoltaics Advanced Solar Energy Conversion}}},\
  Springer series in photonics\ (\bibinfo  {publisher} {Springer-Verlag},\
  \bibinfo {address} {Berlin, Heidelberg},\ \bibinfo {year} {2006})\BibitemShut
  {NoStop}%
\bibitem [{\citenamefont {Toprasertpong}\ \emph {et~al.}(2016)\citenamefont
  {Toprasertpong}, \citenamefont {Fujii}, \citenamefont {Thomas}, \citenamefont
  {Fuehrer}, \citenamefont {Alonso-Alvarez}, \citenamefont {Farrell},
  \citenamefont {Watanabe}, \citenamefont {Okada}, \citenamefont
  {Ekins-Daukes}, \citenamefont {Sugiyama},\ and\ \citenamefont
  {Nakano}}]{toprasertpong:16}%
  \BibitemOpen
  \bibfield  {author} {\bibinfo {author} {\bibfnamefont {K.}~\bibnamefont
  {Toprasertpong}}, \bibinfo {author} {\bibfnamefont {H.}~\bibnamefont
  {Fujii}}, \bibinfo {author} {\bibfnamefont {T.}~\bibnamefont {Thomas}},
  \bibinfo {author} {\bibfnamefont {M.}~\bibnamefont {Fuehrer}}, \bibinfo
  {author} {\bibfnamefont {D.}~\bibnamefont {Alonso-Alvarez}}, \bibinfo
  {author} {\bibfnamefont {D.~J.}\ \bibnamefont {Farrell}}, \bibinfo {author}
  {\bibfnamefont {K.}~\bibnamefont {Watanabe}}, \bibinfo {author}
  {\bibfnamefont {Y.}~\bibnamefont {Okada}}, \bibinfo {author} {\bibfnamefont
  {N.~J.}\ \bibnamefont {Ekins-Daukes}}, \bibinfo {author} {\bibfnamefont
  {M.}~\bibnamefont {Sugiyama}}, \ and\ \bibinfo {author} {\bibfnamefont
  {Y.}~\bibnamefont {Nakano}},\ }\href {\doibase 10.1002/pip.2585} {\bibfield
  {journal} {\bibinfo  {journal} {Prog. Photovolt: Res. Appl.}\ }\textbf
  {\bibinfo {volume} {24}},\ \bibinfo {pages} {533} (\bibinfo {year}
  {2016})}\BibitemShut {NoStop}%
\bibitem [{\citenamefont {Gonzalo}\ \emph {et~al.}(2017)\citenamefont
  {Gonzalo}, \citenamefont {Utrilla}, \citenamefont {Reyes}, \citenamefont
  {Braza}, \citenamefont {Llorens}, \citenamefont {Fuertes~Marr\'on},
  \citenamefont {Al\'en}, \citenamefont {Ben}, \citenamefont {Gonz\'alez},
  \citenamefont {Guzman}, \citenamefont {Hierro},\ and\ \citenamefont
  {Ulloa}}]{gonzalo:17}%
  \BibitemOpen
  \bibfield  {author} {\bibinfo {author} {\bibfnamefont {A.}~\bibnamefont
  {Gonzalo}}, \bibinfo {author} {\bibfnamefont {A.~D.}\ \bibnamefont
  {Utrilla}}, \bibinfo {author} {\bibfnamefont {D.~F.}\ \bibnamefont {Reyes}},
  \bibinfo {author} {\bibfnamefont {V.}~\bibnamefont {Braza}}, \bibinfo
  {author} {\bibfnamefont {J.~M.}\ \bibnamefont {Llorens}}, \bibinfo {author}
  {\bibfnamefont {D.}~\bibnamefont {Fuertes~Marr\'on}}, \bibinfo {author}
  {\bibfnamefont {B.}~\bibnamefont {Al\'en}}, \bibinfo {author} {\bibfnamefont
  {T.}~\bibnamefont {Ben}}, \bibinfo {author} {\bibfnamefont {D.}~\bibnamefont
  {Gonz\'alez}}, \bibinfo {author} {\bibfnamefont {A.}~\bibnamefont {Guzman}},
  \bibinfo {author} {\bibfnamefont {A.}~\bibnamefont {Hierro}}, \ and\ \bibinfo
  {author} {\bibfnamefont {J.~M.}\ \bibnamefont {Ulloa}},\ }\href
  {http://dx.doi.org/10.1038/s41598-017-04321-4} {\bibfield  {journal}
  {\bibinfo  {journal} {Sci. Rep.}\ }\textbf {\bibinfo {volume} {7}},\ \bibinfo
  {pages} {4012} (\bibinfo {year} {2017})}\BibitemShut {NoStop}%
\bibitem [{\citenamefont {Gonzalo}\ \emph {et~al.}(2018)\citenamefont
  {Gonzalo}, \citenamefont {Utrilla}, \citenamefont {Aeberhard}, \citenamefont
  {Llorens}, \citenamefont {Al\'en}, \citenamefont {Fuertes~Marr\'on},
  \citenamefont {Guzman}, \citenamefont {Hierro},\ and\ \citenamefont
  {Ulloa}}]{gonzalo:18_spie}%
  \BibitemOpen
  \bibfield  {author} {\bibinfo {author} {\bibfnamefont {A.}~\bibnamefont
  {Gonzalo}}, \bibinfo {author} {\bibfnamefont {A.~D.}\ \bibnamefont
  {Utrilla}}, \bibinfo {author} {\bibfnamefont {U.}~\bibnamefont {Aeberhard}},
  \bibinfo {author} {\bibfnamefont {J.~M.}\ \bibnamefont {Llorens}}, \bibinfo
  {author} {\bibfnamefont {B.}~\bibnamefont {Al\'en}}, \bibinfo {author}
  {\bibfnamefont {D.}~\bibnamefont {Fuertes~Marr\'on}}, \bibinfo {author}
  {\bibfnamefont {A.}~\bibnamefont {Guzman}}, \bibinfo {author} {\bibfnamefont
  {A.}~\bibnamefont {Hierro}}, \ and\ \bibinfo {author} {\bibfnamefont
  {J.}~\bibnamefont {Ulloa}},\ }in\ \href {\doibase 10.1117/12.2290079} {\emph
  {\bibinfo {booktitle} {Proc. SPIE}}},\ Vol.\ \bibinfo {volume} {10527,
  Physics, Simulation, and Photonic Engineering of Photovoltaic Devices VII,}\
  (\bibinfo {year} {2018})\ pp.\ \bibinfo {pages} {10527}\BibitemShut {NoStop}%
\bibitem [{\citenamefont {Reyes}\ \emph {et~al.}(2012)\citenamefont {Reyes},
  \citenamefont {Gonz{\'a}lez}, \citenamefont {Ulloa}, \citenamefont {Sales},
  \citenamefont {Dominguez}, \citenamefont {Mayoral},\ and\ \citenamefont
  {Hierro}}]{reyes:12}%
  \BibitemOpen
  \bibfield  {author} {\bibinfo {author} {\bibfnamefont {D.~F.}\ \bibnamefont
  {Reyes}}, \bibinfo {author} {\bibfnamefont {D.}~\bibnamefont {Gonz{\'a}lez}},
  \bibinfo {author} {\bibfnamefont {J.~M.}\ \bibnamefont {Ulloa}}, \bibinfo
  {author} {\bibfnamefont {D.~L.}\ \bibnamefont {Sales}}, \bibinfo {author}
  {\bibfnamefont {L.}~\bibnamefont {Dominguez}}, \bibinfo {author}
  {\bibfnamefont {A.}~\bibnamefont {Mayoral}}, \ and\ \bibinfo {author}
  {\bibfnamefont {A.}~\bibnamefont {Hierro}},\ }\href {\doibase
  10.1186/1556-276X-7-653} {\bibfield  {journal} {\bibinfo  {journal}
  {Nanoscale Res. Lett.}\ }\textbf {\bibinfo {volume} {7}},\ \bibinfo {pages}
  {653} (\bibinfo {year} {2012})}\BibitemShut {NoStop}%
\bibitem [{\citenamefont {Utrilla}\ \emph {et~al.}(2014)\citenamefont
  {Utrilla}, \citenamefont {Reyes}, \citenamefont {Ulloa}, \citenamefont
  {Gonz\'alez}, \citenamefont {Ben}, \citenamefont {Guzman},\ and\ \citenamefont
  {Hierro}}]{utrilla:14}%
  \BibitemOpen
  \bibfield  {author} {\bibinfo {author} {\bibfnamefont {A.~D.}\ \bibnamefont
  {Utrilla}}, \bibinfo {author} {\bibfnamefont {D.~F.}\ \bibnamefont {Reyes}},
  \bibinfo {author} {\bibfnamefont {J.~M.}\ \bibnamefont {Ulloa}}, \bibinfo
  {author} {\bibfnamefont {D.}~\bibnamefont {Gonz\'alez}}, \bibinfo {author}
  {\bibfnamefont {T.}~\bibnamefont {Ben}}, \bibinfo {author} {\bibfnamefont
  {A.}~\bibnamefont {Guzman}}, \ and\ \bibinfo {author} {\bibfnamefont
  {A.}~\bibnamefont {Hierro}},\ }\href {\doibase 10.1063/1.4891557} {\bibfield
  {journal} {\bibinfo  {journal} {Appl. Phys. Lett.}\ }\textbf {\bibinfo
  {volume} {105}},\ \bibinfo {pages} {043105} (\bibinfo {year}
  {2014})}\BibitemShut {NoStop}%
\bibitem [{\citenamefont {Lin}\ \emph {et~al.}(2008)\citenamefont {Lin},
  \citenamefont {Ma}, \citenamefont {Chen},\ and\ \citenamefont
  {Lin}}]{lin:08}%
  \BibitemOpen
  \bibfield  {author} {\bibinfo {author} {\bibfnamefont {Y.-T.}\ \bibnamefont
  {Lin}}, \bibinfo {author} {\bibfnamefont {T.-C.}\ \bibnamefont {Ma}},
  \bibinfo {author} {\bibfnamefont {T.-Y.}\ \bibnamefont {Chen}}, \ and\
  \bibinfo {author} {\bibfnamefont {H.-H.}\ \bibnamefont {Lin}},\ }\href
  {\doibase 10.1063/1.3009199} {\bibfield  {journal} {\bibinfo  {journal}
  {Appl. Phys. Lett.}\ }\textbf {\bibinfo {volume} {93}},\ \bibinfo {pages}
  {171914} (\bibinfo {year} {2008})}\BibitemShut {NoStop}%
\bibitem [{\citenamefont {Lin}\ \emph {et~al.}(2013)\citenamefont {Lin},
  \citenamefont {Lin}, \citenamefont {Wang}, \citenamefont {Lin},\ and\
  \citenamefont {Hwang}}]{lin:13}%
  \BibitemOpen
  \bibfield  {author} {\bibinfo {author} {\bibfnamefont {K.-I.}\ \bibnamefont
  {Lin}}, \bibinfo {author} {\bibfnamefont {K.-L.}\ \bibnamefont {Lin}},
  \bibinfo {author} {\bibfnamefont {B.-W.}\ \bibnamefont {Wang}}, \bibinfo
  {author} {\bibfnamefont {H.-H.}\ \bibnamefont {Lin}}, \ and\ \bibinfo
  {author} {\bibfnamefont {J.-S.}\ \bibnamefont {Hwang}},\ }\href
  {http://stacks.iop.org/1882-0786/6/i=12/a=121202} {\bibfield  {journal}
  {\bibinfo  {journal} {Appl. Phys. Express}\ }\textbf {\bibinfo {volume} {6}},\
  \bibinfo {pages} {121202} (\bibinfo {year} {2013})}\BibitemShut {NoStop}%
\bibitem [{\citenamefont {Aeberhard}(2011{\natexlab{a}})}]{ae:jcel_11}%
  \BibitemOpen
  \bibfield  {author} {\bibinfo {author} {\bibfnamefont {U.}~\bibnamefont
  {Aeberhard}},\ }\href {http://dx.doi.org/10.1007/s10825-011-0375-6}
  {\bibfield  {journal} {\bibinfo  {journal} {J. Comput. Electron.}\ }\textbf
  {\bibinfo {volume} {10}},\ \bibinfo {pages} {394} (\bibinfo {year}
  {2011}{\natexlab{a}})}\BibitemShut {NoStop}%
\bibitem [{\citenamefont {Aeberhard}(2018)}]{ae:jmr_18}%
  \BibitemOpen
  \bibfield  {author} {\bibinfo {author} {\bibfnamefont {U.}~\bibnamefont
  {Aeberhard}},\ }\href {\doibase 10.1557/jmr.2017.468} {\bibfield  {journal}
  {\bibinfo  {journal} {J. Mater. Res.}\ }\textbf {\bibinfo {volume} {33}},\
  \bibinfo {pages} {373–386} (\bibinfo {year} {2018})}\BibitemShut {NoStop}%
\bibitem [{\citenamefont {Aeberhard}(2011{\natexlab{b}})}]{ae:nrl_11}%
  \BibitemOpen
  \bibfield  {author} {\bibinfo {author} {\bibfnamefont {U.}~\bibnamefont
  {Aeberhard}},\ }\href {\doibase doi:10.1186/1556-276X-6-242} {\bibfield
  {journal} {\bibinfo  {journal} {Nanoscale Res. Lett.}\ }\textbf {\bibinfo
  {volume} {6}},\ \bibinfo {pages} {242} (\bibinfo {year}
  {2011}{\natexlab{b}})}\BibitemShut {NoStop}%
\bibitem [{\citenamefont {Aeberhard}(2014)}]{ae:jpe_14}%
  \BibitemOpen
  \bibfield  {author} {\bibinfo {author} {\bibfnamefont {U.}~\bibnamefont
  {Aeberhard}},\ }\href {\doibase 10.1117/1.JPE.4.042099} {\bibfield  {journal}
  {\bibinfo  {journal} {J. Photon. Energy}\ }\textbf {\bibinfo {volume} {4}},\
  \bibinfo {pages} {042099} (\bibinfo {year} {2014})}\BibitemShut {NoStop}%
\bibitem [{\citenamefont {W\"urfel}(2009)}]{wuerfel:book_09}%
  \BibitemOpen
  \bibfield  {author} {\bibinfo {author} {\bibfnamefont {P.}~\bibnamefont
  {W\"urfel}},\ }\href
  {http://wwwzb.fz-juelich.de/contentenrichment/inhaltsverzeichnisse/bis2009/ISBN-978-3-527-40857-6.pdf}
  {\emph {\bibinfo {title} {Physics of solar cells : from basic principles to
  advanced concepts}}}\ (\bibinfo  {publisher} {Wiley-VCH},\ \bibinfo {address}
  {Weinheim},\ \bibinfo {year} {2009})\BibitemShut {Stop}%
\bibitem [{\citenamefont {Aeberhard}(2013)}]{ae:mrs_12}%
  \BibitemOpen
  \bibfield  {author} {\bibinfo {author} {\bibfnamefont {U.}~\bibnamefont
  {Aeberhard}},\ }\href {\doibase 10.1557/opl.2013.226} {\bibfield  {journal}
  {\bibinfo  {journal} {MRS Proceedings}\ }\textbf {\bibinfo {volume} {1493}},\
  \bibinfo {pages} {91} (\bibinfo {year} {2013})}\BibitemShut {NoStop}%
\bibitem [{Note1()}]{Note1}%
  \BibitemOpen
  \bibinfo {note} {The qualitative extraction picture does not depend on the
  exact value of $\tau $, as long as 1/$\tau $ is comparable to the rate of
  carrier escape to the contact reservoirs.}\BibitemShut {Stop}%
\bibitem [{\citenamefont {Berbezier}\ and\ \citenamefont
  {Aeberhard}(2015)}]{berbezier:15}%
  \BibitemOpen
  \bibfield  {author} {\bibinfo {author} {\bibfnamefont {A.}~\bibnamefont
  {Berbezier}}\ and\ \bibinfo {author} {\bibfnamefont {U.}~\bibnamefont
  {Aeberhard}},\ }\href {\doibase 10.1103/PhysRevApplied.4.044008} {\bibfield
  {journal} {\bibinfo  {journal} {Phys. Rev. Applied}\ }\textbf {\bibinfo
  {volume} {4}},\ \bibinfo {pages} {044008} (\bibinfo {year}
  {2015})}\BibitemShut {NoStop}%
\bibitem [{\citenamefont {Aeberhard}(2016)}]{ae:apl_16}%
  \BibitemOpen
  \bibfield  {author} {\bibinfo {author} {\bibfnamefont {U.}~\bibnamefont
  {Aeberhard}},\ }\href {\doibase http://dx.doi.org/10.1063/1.4959244}
  {\bibfield  {journal} {\bibinfo  {journal} {Appl. Phys. Lett.}\ }\textbf
  {\bibinfo {volume} {109}},\ \bibinfo {eid} {033906} (\bibinfo {year}
  {2016})}\BibitemShut {NoStop}%
\end{thebibliography}
\end{document}